\documentclass[sigconf]{acmart}
\AtBeginDocument{%
  }

\setcopyright{none}
\usepackage{amsmath}
\usepackage{algorithm}
\usepackage{algpseudocode}
\begin{document}

\title{Comparing Domain-Model Similarity Metrics Against Human Expert Ratings}

\author{Vasiliy Seibert}
\affiliation{%
  \institution{Institute for Software and Systems Engineering, TU Clausthal}
  \city{Clausthal-Zellerfeld}
  \state{Lower Saxony}
  \country{Germany}
}
\email{vasiliy.seibert@tu-clausthal.de}
\orcid{0000-0002-7121-6816}

\begin{abstract}
Domain models are a primary artefact in model-driven software engineering, where they capture the shared understanding between stakeholders and serve as the contractual basis for downstream software development. Automatic comparison of these semantic models has diverse application areas such as requirements engineering, education, automatic generation of domain models and model reuse and repository mining. The literature offers a variety of presented metrics, but for practitioners there is no defensible way to choose between them. The contribution of this paper is the implementation of five such metrics, their execution on a fixed set of 39 domain-model comparisons and the comparison of each metric's output against the human expert ratings produced for the same comparisons. Two research questions are addressed. RQ1 asks how close, on average, each metric is to the human expert rating across the 39 comparisons. RQ2 asks how consistent each metric's per-comparison distance from the human expert rating is. The findings reveal that no single metric achieves dominance across all criteria; rather, different metrics each yield competitive results on individual criteria — some closest on average, others best preserving the per-pair ordering — which suggests that an ensemble approach combining multiple metrics may serve as a viable substitute for human expert grading. The metric implementations are an artefact of this work and are published in accordance with the FAIR4RS recommendations (https://doi.org/10.5281/zenodo.20942596).
\end{abstract}



\received{2026-06-13}

\maketitle


\section{Introduction}
\label{sec:1_introduction}


Domain models are a primary artifact in model-driven software engineering (MDSE)~\cite{chen2023automated, chen2023gpt4goalmodeling, yuan2020structural}. A domain model captures the entities, attributes, and associations of an application domain in a notation that is read by stakeholders from different backgrounds \textemdash{} domain experts, requirements engineers, software architects, and developers \textemdash{} and serves as a basis for downstream software development. The diagram's role is therefore twofold: it is the artefact on which all participants converge to gain a mutual understanding of the system to be built, and it is the input to subsequent design, code generation, and verification activities. \newline


A well-defined, deterministic metric for comparing domain models would benefit multiple areas. In automated generation of domain models~\cite{chen2023automated, chen2023gpt4goalmodeling, yang2024multistep, yang2022extracting}, the metric is the measurement of a generation approach's performance against a reference. In education~\cite{modi2021tool, boubekeur2020automatic, bouali2025llmgrading, singh2022detecting}, the metric could automate the grading of student-submitted class diagrams against an instructor's reference and shorten the feedback loop. In requirements engineering~\cite{cech2019matching, yuan2020structural}, the metric could let a modeler compare newly created models with models from other projects.  In model reuse and repository mining~\cite{song2024deep, cech2019matching, yuan2020structural, triandini2021automated}, the metric is the basis on which a repository query ranks candidates, a design-pattern detector flags matches, and a reuse decision is made. Without a well-defined, deterministic metric, all of these activities reduce to manual inspection. \newline


The metrics that have been proposed in the literature are diverse in approach. One approach is rule-based 
mistake detection~\cite{modi2021tool, singh2022detecting}: the metric enumerates a list of required-model 
properties (e.g., expected classes, expected attributes, expected associations) and reports the missing or 
extra elements as a mistake list, scored at the level of the mistake type. Another approach uses graph edit 
distance (GED) on attributed relational graphs~\cite{cech2019matching}: the metric reduces the similarity 
between two models to the cost of the cheapest sequence of edit operations that transforms one graph into the 
other, and the similarity is reported as a normalised score derived from that cost. Yet another approach is 
structural matching on a domain-specific graph representation, exemplified by the UML Class Graph 
(UCG)~\cite{yuan2020structural}: the metric decomposes the similarity into an intra-cluster component (the 
quality of grouping related elements within one model) and an inter-cluster component (the alignment of 
clusters across the two models). A further approach is a semantic+structural 
pipeline~\cite{triandini2021automated}: the metric separates the structural similarity of the two models 
(computed on the diagram's structure) from the semantic similarity (computed on the lexical information in the 
class names) and combines the two. One more approach is \mbox{deep-learning-based} similarity, exemplified by 
the SimGNN approach to UML use case models~\cite{song2024deep}, in which the candidate and reference models 
are encoded as graphs and a graph neural network produces the similarity score. \newline


The diversity of the proposed metrics, combined with the fact that each is applied on its own dataset means that the field has many candidate metrics but no defensible way to choose between them.\newline


So that the field has arguments grounded in a common dataset for choosing one similarity metric over another, this paper implements five metrics from the literature~\cite{singh2022detecting, cech2019matching, yuan2020structural, triandini2021automated} and compares their outputs against human expert ratings on a fixed set of 39 domain-model comparisons~\cite{chen2023automated}. Two research questions are formulated before the comparison. \textbf{RQ1} is formulated as follows: \emph{how close, on average, is each metric to the human expert rating across the 39 comparisons?} The question matters because alignment with the human judgement on average is an indicator that the metric captures the notion of similarity that a human grader would apply, and is the operational form of a metric's central tendency. \textbf{RQ2} is formulated as follows: \emph{how consistent is each metric's per-comparison distance from the human expert rating?} The question matters because a metric that systematically undervalues or overvalues similarity can still be useful \textemdash{} what matters for many practical applications is that the metric preserves the ordering of similarity across pairs, so that two diagrams judged more similar by the human are also judged more similar by the metric; the per-comparison consistency is the operational form of a metric's reliability as a ranker.\newline


Two risks attend this approach. The first is the risk of an incorrect implementation: the metrics in the literature are described at the level of a published paper, and a faithful re-implementation requires a careful reading of the source, which is not always unambiguous; an implementation error in any of the 5 metrics would invalidate the cross-metric comparison for that metric. The second is the risk of an unfair translation: the 5 metrics emit scores in different units, and the comparison requires each score to be translated into a per-element F1 (class, attribute, association) in the same scale as the human expert rating, a translation that is per-metric and that risks putting some metrics on a scale they were not designed to operate in.\newline


To address the first risk, a specification is derived from the publication of each metric, two different large language models implement the same specification, and the two implementations are tested for deterministic agreement: if both implementations produce the same result, the specification is assumed to be well-defined and the implementations are assumed to be correct, with the specification serving as the explicit record of the metric as understood in this work. To address the second risk, the specification is extended to translate the metric's output into the per-element F1 scale of the human expert rating, and the translation is tested for partial-order preservation against the metric's native output, so that the ordering of the metric results is consistent with the ordering of the original output across the 39 comparisons.\newline


To encourage reuse, the metric implementations are a supplementary research artifact to this work in accordance with the FAIR4RS recommendations~\cite{chuehong2022fair4rs} \url{https://doi.org/10.5281/zenodo.20942596} \newline


\section{Background and Related Work}
\label{sec:2_background_relatedwork}

\subsection{Related Work}

Education is one application area that would benefit from an established Domain Model Similarity Metric. Modi, Taher, and Mahmud~\cite{modi2021tool} develops a Java-based tool that reads student and instructor solution diagrams as input and produces an automatic grade for the UML diagram types submitted in software-engineering and system-analysis courses. Boubekeur, Mussbacher, and McIntosh~\cite{boubekeur2020automatic} combine a simple heuristic with machine-learning techniques to predict approximate letter grades. Bouali et al.~\cite{bouali2025llmgrading} compares LLM-generated scores against three teaching assistants on 92 student submissions of UML class diagrams in a software-design course. Singh, Boubekeur, and Mussbacher~\cite{singh2022detecting} propose a Mistake Detection System (MDS) that is a rule-based system which automatically indicates the exact location and type of the mistake in a student solution when compared with a correct solution, covering 83 out of 97 identified different types of mistakes that may exist in a student solution. On real student solutions, the authors report that, when synonyms are considered by MDS, the system achieves a recall of 0.93 and a precision of 0.79. The metric is evaluated on its own real-student dataset and is not cross-compared with other model-similarity metrics in the paper. Fauzan, Siahaan, Rochimah, and Triandini~\cite{fauzan2021automated} decomposes the assessment into a structural similarity, calculated using the diagram's structure while ignoring its lexical information, and a semantic similarity, calculated using the lexical information in the class diagram. The authors report that experts see semantic and structural similarities equally during assessment, and that the proposed approach shows agreement with experts in class-diagram similarity assessment. The two metrics are validated against expert agreement on the authors' own student-diagram dataset. \newline


General model-driven engineering and requirements engineering are additional application areas. Chen, Yang, Chen, Hern\'andez L\'opez, Mussbacher, and Varr\'o~\cite{chen2023automated} conduct a comparative study of using large language models for fully automated domain modeling on a dataset of examples with reference solutions created by modeling experts, framing domain modeling as a core software-engineering activity. Chen, Chen, Varr\'o, and Mussbacher~\cite{chen2023gpt4goalmodeling} explore the use of GPT-4 for creating goal-oriented models in the Goal-oriented Requirement Language (GRL), positioning their work as a requirements-engineering application of LLM-based modeling. Yang, Chen, Chen, Mussbacher, and Varr\'o~\cite{yang2024multistep} extend the same line of LLM-based domain modeling with a multi-step iterative framework and a self-reflection mechanism, again targeting general software-engineering modeling. Yang and Sahraoui~\cite{yang2022extracting} target model-driven engineering by proposing an automated approach for extracting UML class diagrams from natural-language software specifications. \newline


Model reuse and model-repository mining are a further application area. Song, Wang, Wang, Lin, and Hu~\cite{song2024deep} list model reuse, model validation, model retrieval, and pattern detection as concrete use cases for similarity calculation of UML models, and they focus their work on use case models, for which they observe that prior similarity-calculation research has been limited; they transform UML use case models into use case graphs and then adopt the Similarity Graph Neural Network (SimGNN) to calculate the structural similarity between the resulting graphs, while semantic similarity is calculated separately with Term Frequency-Inverse Document Frequency (TF-IDF) and cosine similarity, and the two results are then combined; the authors report that their approach is evaluated by comparison to two baselines, namely cosine similarity combined with TF-IDF and cosine similarity combined with edit distance, and that it achieves higher accuracy and lower runtime than both. \v{C}ech~\cite{cech2019matching} introduces a class-model distance computation framework that can be used for comparing class models in model repositories, with application to both pairwise model comparison and design-pattern detection. Yuan, Yan, and Ma~\cite{yuan2020structural} represent a UML class diagram as a graph called UCG and decompose the structural similarity of a class diagram into two aspects: intra-structure, which refers to the composition of each class, and inter-structure, which refers to the relationships between classes. For inter-structure similarity they propose an algorithm based on UCG Maximum Common Subgraph Sequence, and for intra-structure similarity they introduce UCG edit distance. The authors report experimental results both within a domain and across domains, but do not cross-compare their structural similarity measure with prior semantic or structural UML similarity metrics. \newline


\section{Comparison Methodology}
\label{sec:3_comparison_methodology}

\subsection{Automated Domain Modeling with Large Language Models}

Chen, Yang, Chen, Hern\'andez L\'opez, Mussbacher, and Varr\'o~\cite{chen2023automated} introduce a study in which they compare LLM-generated against expert-authored reference models. This paper uses this comparison as the basis for the literature-metric comparison in this work.\newline
The reference models span ten application domains: a bus transportation management system (BTMS), a second-hand delivery and pickup system (H2S), a medical laboratory test management system (LabTracker), a celebrations and order tracking system (CelO), a team sports management system (TeamSports), a smart home automation system (SHAS), a university OTS system, two board games (Block, Tile-O), and a hotel booking management system (HBMS). The Reference Models are mostly expressed as images.\newline
The authors apply \emph{Davinci} (text-davinci-003, the GPT-3.5 completion model) \cite{openai2023gpt35docs}, \emph{Turbo} (gpt-3.5-turbo) \cite{ouyang2022training}, and \emph{GPT-4} \cite{openai2023gpt4} using five prompt settings: \emph{0-shot} (task instruction plus target description only), \emph{1-shot-btms} (one BTMS exemplar), \emph{1-shot-h2s} (one H2S exemplar), \emph{2-shot} (BTMS and H2S combined as exemplars), and \emph{\mbox{chain-of-thought}} (CoT, \mbox{sentence-by-sentence} reasoning over H2S).\newline


\noindent\textbf{Generated Domain Model} The LLM Generated Domain Models describe Enumerations, Classes and Relationships according to the following structure:

\begin{small}
\begin{verbatim}
Enumerations:
(...)
TestResult(NEGATIVE, POSITIVE)

Classes:
(...)
Requisition(validFromDate: Date)
Test(group: String, duration: Integer)
(...)

Relationships:
1 Doctor associate 0..* Requisition
1 Patient associate 0..* Requisition
1 Requisition contain 1..* Test
1 Test associate 0..1 Appointment
1 Lab associate 0..* Appointment

Inheritance:
None

Composition:
(...)
1 Test contain 1 TestOutcome(result: TestResult, (...))
\end{verbatim}
\end{small}


The eight models (ten examples minus BTMS and H2S, which appear in the prompts and are therefore excluded) contain in total $135$ classes, $221$ attributes, and $177$ relationships, with the ten-example totals being $155$ classes, $250$ attributes, and $204$ relationships. Crossed with the three LLMs and the five prompt settings, the paper therefore contains $3 \times 8 \times 5 = 120$ \mbox{generated-vs-reference} comparisons. \newline


The grading itself is performed by human raters in a two-round consensus process. Each generated model is graded against the corresponding reference model per element type — \emph{class}, \emph{attribute}, \emph{relationship} by applying four categories (c1, c2, c3 and c4) distinguishing exact matches, semantically equivalent matches, partial matches, and elements with no match in the reference model. Category $c_1$ captures generated elements that have a direct semantic counterpart of the same type in the reference model, e.g., the \texttt{Person} class of the reference model, which is matched against the \texttt{User} class of the generated model because both capture the same domain concept. Category $c_2$ captures generated elements that match an element of a different type in the reference model but are semantically equivalent to that element, e.g., the \texttt{SecondHandArticle} class, which the reference model represents with a boolean attribute \texttt{discarded} and the generated model represents with an enumeration \texttt{Status} taking the two literals \texttt{AVAILABLE} and \texttt{DISCARDED}; the two representations are semantically equivalent but differ in type. Category $c_3$ captures elements that only partially match a reference element, e.g., the \texttt{0..1 Route associate * Item} relationship of the reference model, which is partially matched when the generated model marks \texttt{Route} as non-optional (\texttt{1 Route associate * Item}, a multiplicity mistake). Category $c_4$ captures generated elements that have no match in the reference model, e.g., the abstract \texttt{UserRole} class of the reference model, which has no equivalent modelling element in the generated model. \newline


\begin{equation*}
S(x) = \begin{cases} 1 & \text{if } x \in c1 \text{ or } c2, \\ 0.5 & \text{if } x \in c3, \\ 0 & \text{if } x \in c4, \end{cases}
\end{equation*}\newline


To turn the four categories into a numeric score per element, the authors define a scoring function $S(x)$ that awards $1$ point to generated elements in $c_1$ or $c_2$ (direct or semantically equivalent match), $0.5$ points to elements in $c_3$ (partial match), and $0$ points to elements in $c_4$ (no match). For each modelling element type, the precision and recall are then computed by summing $S(x)$ over the generated and the reference elements separately, and dividing by the respective set sizes. For classes with $m$ generated elements and $n$ reference elements,

\begin{equation}
\text{Precision}_{C} = \frac{\sum_{i=1}^{i=m} S(C_i)}{m},
\label{eq:chenPrecision}
\end{equation}\newline

\begin{equation}
\text{Recall}_{C} = \frac{\sum_{i=1}^{i=n} S(C_i)}{n},
\label{eq:chenRecall}
\end{equation}\newline


\begin{equation}
\text{F1}_C = \frac{2 \times \text{Precision}_C \times \text{Recall}_C}{\text{Precision}_C + \text{Recall}_C}.
\label{eq:chenF1}
\end{equation}\newline

and the same definitions apply with $A$ for attributes and $R$ for relationships (Equations 1--4 of \cite{chen2023automated}). The procedure is defined three times rather than once over an aggregate score, and the output of the grading is a $3$-tuple of F1 values, $\langle \text{F1}_C, \text{F1}_A, \text{F1}_R \rangle$, one per element type. This $3$-tuple is the de-facto output of the manual grading, and it is the unit of comparison that this work reuses as the ground truth.\newline

\subsection{Reusing Human Expert Ratings}

In order to reuse the human expert ratings and test automated metrics against them, we must translate the data into a format that can be reliably processed by a parser. The Chen et al. paper distributes its models only as rendered images and a CSV of judgements. We therefore translated the 8 reference models and 39 generated models into PlantUML~\cite{plantuml}, producing a dataset of 39 PlantUML comparisons. Each comparison pairs a reference model with a generated model and carries the corresponding human F1 3-tuple $\langle \text{F1}_C, 	ext{F1}_A, 	ext{F1}_R 
angle$ from \S3.1. \newline


\noindent\textbf{Translated Generated Domain Model to PlantUML} The LLM Generated Domain Models were manually translated to PlantUML:

\begin{small}
\begin{verbatim}
@startuml
enum RepetitionInterval {
    WEEKLY
    MONTHLY
    HALF_YEARLY
    YEARLY
}
class Requisition {
    validFromDate : Date
}
class Test {
    group : String
    duration : Integer
}

Doctor "1" -- "0..*" Requisition 
Patient "1" -- "0..*" Requisition 
Requisition "1" *-- "1..*" Test 
Test "1" -- "0..1" Appointment 
Lab "1" -- "0..*" Appointment 

Test "1" *-- "1" TestOutcome 

@enduml
\end{verbatim}
\end{small}


As a consequence, every candidate metric in this study must accept as input two PlantUML domain-model strings — one reference, one generated — and output a 3-tuple of similarity scores $\langle s_C, s_A, s_R 
angle$, one per modelling element type, with each $s \in [0, 1]$. The three scores are aligned with the human categories.


\begin{small}
\begin{verbatim}
Metric M
  compute(ReferencePlantUML, GeneratedPlantUML)
    -> (class_score      : [0, 1],
       attribute_score   : [0, 1],
       association_score : [0, 1])
\end{verbatim}
\end{small}


\section{Implementation Methodology}
\label{sec:4_implementation_methodology}

\begin{figure*}[htbp]
\centering
\includegraphics[width=0.75\textwidth]{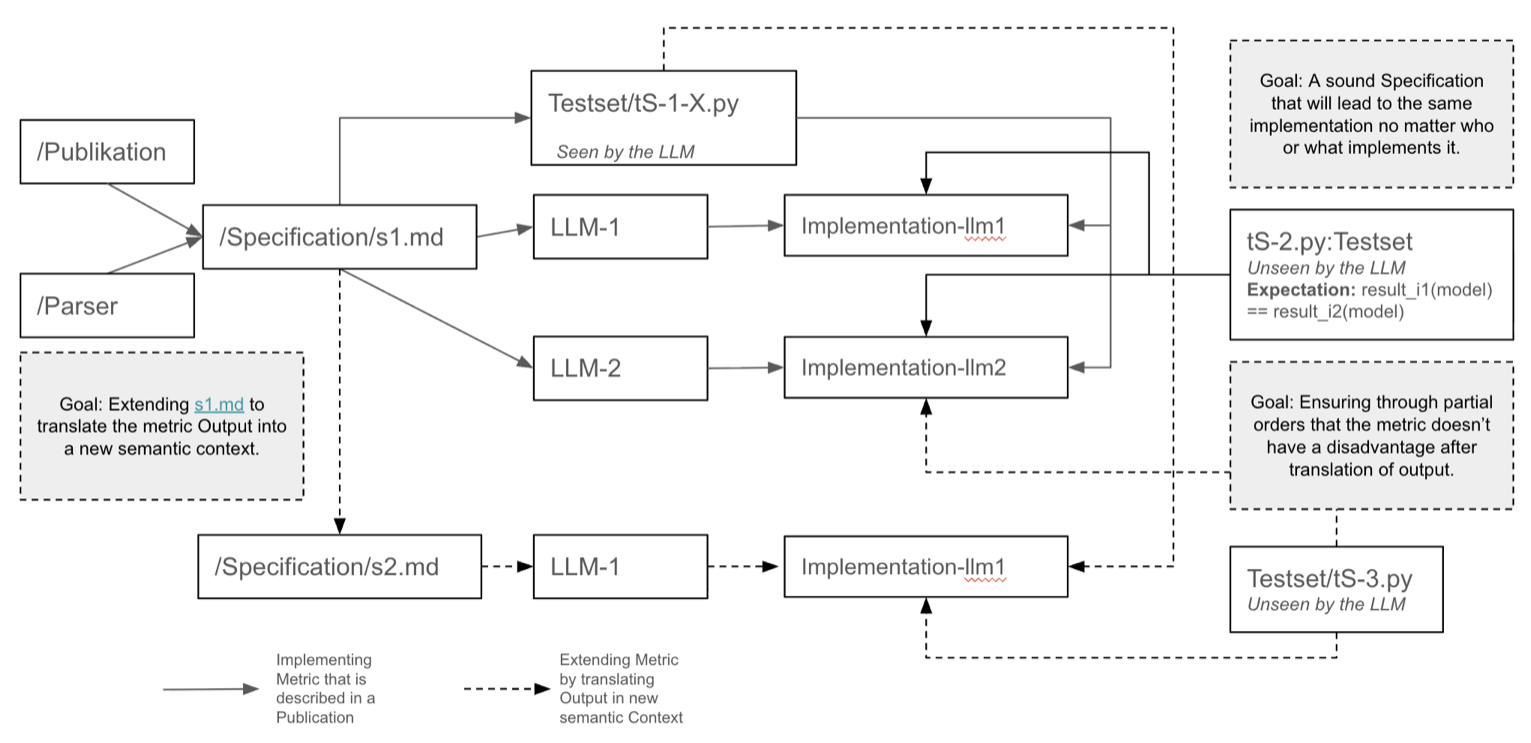}
\caption{Spec → invariants → @icontract → code pattern.}
\label{fig:4-1}
\end{figure*}


An incorrect implementation of any of the five metrics would confound the comparison, because disagreement between a metric and the human expert rating could then be attributed to the implementation rather than to the metric itself, and the comparison's validity rests on every implementation being a faithful restatement of its source paper.\newline


To address this risk, each metric is first restated as a specification before any code is written. The specification is expressed in a markdown file and Python Scripts containing docstrings, which docompose the metric into its functional-logical components and declares, for each component, the preconditions that its inputs must satisfy and the post-conditions that its output must satisfy, in the style of design by contract~\cite{meyer1992applying}. The specification is then handed, unchanged, to two independent large language models, each of which produces a separate Python implementation inside the predefined Python Scripts. The two implementations are executed on 39 domain-model comparisons that \S{}3 describes, and they are required to produce identical output on every comparison; if they do, the specification is assumed to be unambiguous and the implementations are assumed to be a correct implementation. The specification, not the code, is the artifact that the comparison is based on: it is the explicit record of the metric as understood in this work, and every assumption or deviation that the implementation makes is written into it so that the comparison can be audited against the source paper. The second stage of the method extends the specification to translate the metric's native output into the per-element (class, attribute, association) scale of the human expert rating, so that the five metrics become comparable both to each other and to the human judgement. \newline


The structure of a specification is illustrated by the following example, which restates the rule-based mistake detector of Singh, Boubekeur and Mussbacher~\cite{singh2022detecting}:
\begin{small}
\begin{verbatim}
metric(instructor_model, student_model):
   mapping |= mapClasses(instructor_model, student_model)
   mapping |= mapRelations((...))
   mistakes  = checkClasses(mapping)
   mistakes += checkRelations(mapping)
   mistakes += checkMissing(mapping)
   return mistakes
   requires:
       isValidModel(instructor_model),
       isValidModel(student_model)
   ensures:
       isValidMistakes(mistakes)
(...)

isValidModel(M):
    checks if model adheres to the data structure.
    class names are unique within M
    enumeration names are unique within M
    every relation references classes that exist in M
    every attribute belongs to exactly one class in M

(...)
\end{verbatim}
\end{small}
The example opens with a top-level metric function that expects two models as input, and both inputs must adhere to the isValidModel precondition; the function returns a list of mistakes, which must adhere to the isValidMistakes postcondition. The pre- and post-conditions are defined beforehand by the authors against the source paper, not by the LLM, so that the contract is fixed before any implementation is produced. Beneath the top-level function, the example decomposes the metric into five sub-functions \textemdash{} mapClasses, mapRelations, checkClasses, checkRelations, checkMissing \textemdash{} each with its own requires/ensures pair, so that the contract is layered: the post-condition of one function becomes the precondition of the next. The Invariants (e.g. isValidModel, isValidMapping, and isValidMistakes) are defined as @icontract.require or @icontract.ensure ~\cite{icontract} decorators in the Python Scripts, where they are checked at runtime on every function call. The specification is therefore not only documentation but a contract that the code is obliged to satisfy. \newline


The two independent implementations of each specification are produced by two open source large language models from different families, \cite{kimi-k2.7-code} and \cite{glm-5.2}, chosen so that agreement between the two is not an artefact of a single model agreeing with itself. Both models are driven through the same agentic harness, \texttt{opencode}, which gives each model equivalent access to the specification file, the shared PlantUML parser, and the test directory, so that any difference in the produced implementation is attributable to the model's interpretation of the specification rather than to differential tooling. Each implementation is produced in a fresh session from the same specification prompt, and the two sessions do not see each other's code, which is the independence that the agreement test in the next paragraph relies on. The role of the large language models is deliberately bounded: the models implement the specification, they do not author it, and they do not judge whether the two implementations agree; the specification is written by the authors against the source paper, and the agreement check is a deterministic execution of both implementations on the same 39 inputs. \newline


The first test procedure, tS-1, follows a \mbox{test-driven-development} approach: the pre- and post-conditions are exposed to the large language model in the specification, and are embedded in the Python skeleton files as @icontract.require and @icontract.ensure decorators, so that they are executed on every function call; a contract violation raises an explicit runtime error rather than degrading silently. The second test procedure, tS-2, checks the two redundant implementations for determinism: both are executed on the 39 domain-model comparisons, and if they produce identical output on every pair, the specification is taken to be unambiguous; if they disagree, the specification is amended and the implementations are regenerated. The third test procedure, tS-3, is applied after the specification is extended to the second stage, where the metric's native output is projected into the per-element (class, attribute, association) scale, and it verifies that the projection preserves the partial ordering of the native output, so that the relative ordering of model pairs is consistent between the first-stage output and the projected output. \newline

\subsection{Literature Metrics}

Five candidate metrics are derived from four source papers. Metrik-1 (M-1) is based on the rule-based mistake detector of Singh, Boubekeur and Mussbacher~\cite{singh2022detecting}. The paper describes 97 distinct mistake types but reports that the presented rules cover only 83 of the 97 types; the remaining 14 types are flagged as out of scope. The native output of M-1 is a mistake list : a deduplicated, sorted list of detected mistakes. To extend the Metric, the mistake list is projected into the required triple of class-score, attribute-score, association-score. The projection works by bucketing the detected mistakes into three groups by their mistake identifiers and normalising each count by the corresponding element count of the instructor model, so that each score is 1. \newline


Metrik-2 (M-2) is based on the \mbox{graph-edit-distance} metric proposed by \v{C}ech~\cite{cech2019matching} for attributed relational graphs. The method reduces the similarity between two models to the cost of the cheapest sequence of edit operations (insert node, delete node, insert edge, delete edge, substitute attribute) that transforms one graph into the other, where the cost weights are paper-specific. The native output of M-2 is a single normalised similarity scalar in 0,1. To extend the Metric, the scalar is projected into the required triple of class-score, attribute-score, association-score. The projection decomposes the aggregate edit cost into three components: vertex-level edit operations produce class-score, edge-level edit operations produce association-score, and attribute-level distance is computed as a Hungarian matching over the attribute sets of each mapped vertex pair to produce attribute-score, with each component normalised by the appropriate maximum cost so that each score is in 0,1. \newline


Metrik-3 (M-3) is based on the UCG structural-similarity metric of Yuan, Yan and Ma~\cite{yuan2020structural}. Each input model is first transformed into a UML Class Graph (UCG): classes become vertices, attributes become attribute vertices linked by attribute edges, and relationships become typed relationship edges whose tag encodes the relationship type (association, inheritance, aggregation, composition, dependency). A backtracking algorithm then finds all maximum-cardinality sets of relationship edges in one graph that can be matched injectively to relationship edges in the other with the same tag and a consistent vertex mapping; this is the UML Maximum Common Subgraph (UMCS). From the UMCS, the inter-structure similarity is the size of the common subgraph normalised by the minimum number of relationship edges in either graph, and the intra-structure similarity is computed by comparing the attribute sets of each matched class-vertex pair via edit distance. The native output of M-3 is the weighted combination of the two, with the inter-structure component at $\theta = 0.9$ and the intra-structure component at $0.1$. To extend the Metric, the combined score is projected into the required triple of class-score, attribute-score, association-score. The projection averages three structural sub-scores with three required F1 scores.\newline


Metrik-4 (M-4) and Metrik-5 (M-5) are both based on the semantic-structural similarity metric of Triandini~\cite{triandini2021automated}. The metric separates a semantic component (computed from class-name and relationship similarity via WordNet-based cosine similarity) from a structural component (computed from intra-cluster and inter-cluster \mbox{graph-edit-distance} similarity on a UCG representation), and combines the two with equal weighting. The native output is a 7-field result comprising the overall similarity, the semantic and structural components, and four sub-scores: propSim (property similarity), relSim (relationship similarity), intraSim (intra-structure similarity), and interSim (inter-structure similarity). To extend the Metric, the 7-field result is projected into the required triple of class-score, attribute-score, association-score. The two metrics share the same source paper and the same native output but differ in the projection step, because the paper's description of how the model-level similarity is mapped into the triple admits two plausible readings: in M-4, the semantic sub-score maps to class-score, the property similarity to attribute-score, and the relationship similarity to association-score; in M-5, the intra-structure similarity maps to class-score and attribute-score, and the inter-structure similarity maps to association-score. \newline


Although M-3 and M-5 both operate on the same UCG representation, the two projections pair classes differently. M-3 pairs classes only as a by-product of its UMCS relationship-edge alignment: two classes are paired only because they appear as endpoints of matched relationship edges. M-5, by contrast, pairs classes directly via a greedy \mbox{graph-edit-distance} over per-class attribute subgraphs, so two classes are paired because their attribute vertices and attribute edges are structurally similar regardless of any relationship context. The two projections therefore surface different per-pair mistakes on the same 39 inputs, which is the practical reason for keeping both M-3 and M-5 in the comparison rather than collapsing them into a single metric.\newline


\section{Quantitative Results}
\label{sec:5_quantitative_results}

\subsection{The Consistency Table}

For a fixed metric $m$ and a fixed element type $e \in \{\text{class},\ \text{attr},\ \text{rel}\}$, the consistency table reports four statistics computed from the $N = 39$ per-pair signed deltas $d_i^{(e)} = m_i^{(e)} - h_i^{(e)}$, where $m_i^{(e)}$ is the metric's per-element score on pair $i$ and $h_i^{(e)}$ is the human per-element F1 on pair $i$. The four quantities are:

\begin{equation*}
\mathrm{MAD}_e \;=\; \frac{1}{N}\sum_{i=1}^{N} \bigl|d_i^{(e)}\bigr|
\end{equation*}

The MAD is the mean absolute difference between the metric's per-element score and the human F1, so a smaller MAD indicates that the metric is closer to the human F1 on average for that element type.

\begin{equation*}
|\mathrm{bias}|_e \;=\; \biggl|\frac{1}{N}\sum_{i=1}^{N} d_i^{(e)}\biggr|
\end{equation*}

The $|\mathrm{bias}|$ is the absolute value of the mean signed delta and captures a constant offset between the metric's scores and the human F1: a non-zero $|\mathrm{bias}|$ with a small residual std means the metric is consistently off by a roughly constant amount (calibratable by a shift), and a near-zero $|\mathrm{bias}|$ means the metric is centred on the human F1 on average.

\begin{equation*}
\mathrm{residual\_std}_e \;=\; \sqrt{\frac{1}{N-1}\sum_{i=1}^{N}\bigl(d_i^{(e)} - \mathrm{bias}_e\bigr)^2}
\end{equation*}

The residual std is the sample standard deviation of the per-pair signed deltas around their mean (with the $1/(N-1)$ denominator), expressed on the same scale as the MAD; a smaller residual std indicates that the per-pair distance is tightly distributed around its mean, while a larger residual std indicates that the per-pair distance varies across pairs.

\begin{equation*}
r_e \;=\;
\frac{\sum_{i=1}^{N}\bigl(m_i^{(e)} - \bar m^{(e)}\bigr)\bigl(h_i^{(e)} - \bar h^{(e)}\bigr)}
{\sqrt{\sum_{i=1}^{N}\bigl(m_i^{(e)} - \bar m^{(e)}\bigr)^2}\;\sqrt{\sum_{i=1}^{N}\bigl(h_i^{(e)} - \bar h^{(e)}\bigr)^2}}
\end{equation*}

The Pearson $r$ is the linear correlation between the metric's per-element score and the human per-element F1, so a high $r$ means the metric preserves the per-pair ordering of the human F1, while a near-zero $r$ means the metric's per-pair score is statistically independent of the human's. \newline


The MAD, residual std, bias, and Pearson r for each metric, on each of the 3 element types, are shown in Table~\ref{tab:consistency}.\newline

\begin{table*}[htbp]
\centering
\centering\footnotesize
\caption{Per-element MAD, residual std, $|$bias$|$, and Pearson r for the 5 candidate metrics on the 39 pairs. Best value per column in \textbf{bold}; lower is better for MAD, r.std, and $|$bias$|$; higher is better for r.}
\label{tab:consistency}
\begin{tabular}{lrrrrrrrrrrrr}
\toprule
Metric & Class MAD & Class r.std & Class $\|$bias$\|$ & Class r & Attr MAD & Attr r.std & Attr $\|$bias$\|$ & Attr r & Rel MAD & Rel r.std & Rel $\|$bias$\|$ & Rel r \\
\midrule
M-1 & 0.15 & 0.09 & 0.14 & 0.37 & 0.23 & 0.19 & 0.19 & 0.23 & \textbf{0.13} & 0.15 & 0.08 & -0.13 \\
M-2 & 0.19 & 0.12 & 0.18 & 0.30 & 0.15 & 0.17 & \textbf{0.06} & 0.18 & 0.18 & 0.13 & 0.17 & 0.37 \\
M-3 & 0.17 & 0.09 & 0.17 & 0.36 & 0.21 & \textbf{0.12} & 0.21 & \textbf{0.65} & \textbf{0.13} & 0.16 & \textbf{0.00} & 0.05 \\
M-4 & 0.09 & \textbf{0.07} & 0.07 & \textbf{0.42} & \textbf{0.14} & 0.15 & 0.09 & 0.32 & 0.27 & \textbf{0.11} & 0.27 & \textbf{0.42} \\
M-5 & \textbf{0.07} & 0.09 & \textbf{0.03} & 0.18 & 0.15 & 0.15 & 0.10 & 0.38 & 0.26 & \textbf{0.11} & 0.26 & 0.20 \\
\bottomrule
\end{tabular}

\end{table*}


The class scores express how similar the two models are at the level of their classes \textemdash{} the entities that populate the domain. Across the 5 metrics, the class MAD spreads from 0.07 to 0.19. M-5 is the closest to the human F1 on average, with a MAD of 0.07, a bias of 0.03, and a residual std of 0.09. M-4 is a close second on MAD at 0.09, and has the tightest spread around its mean with a residual std of 0.07, as well as the highest Pearson r in the column at 0.42, meaning that M-4 preserves the per-pair ordering of the human F1 better than any other metric on the class element. M-1 sits in the middle: a MAD of 0.15, a residual std of 0.09, a bias of 0.14, and a Pearson r of 0.37, meaning it is moderately close on average but carries a constant underestimation of 0.14. M-3 is similar to M-1 on MAD at 0.17 and residual std at 0.09, but carries a larger bias of 0.17 and a slightly lower r of 0.36. M-2 is the weakest on class: the highest MAD at 0.19, the highest residual std at 0.12, the highest bias at 0.18, and the second-lowest r at 0.30, meaning that it is both far from the human on average and a poor ranker. The class scores admit no single dominant metric: the metric with the lowest MAD, M-5 at 0.07, is not the metric with the lowest residual std, M-4 at 0.07, or the highest r, M-4 at 0.42. A practitioner who wants the metric whose class score is closest to the human on average picks M-5; a practitioner who wants the metric whose class score preserves the human's per-pair ranking picks M-4.\newline


The attribute scores express how similar the two models are at the level of their attributes \textemdash{} the properties that describe each class. Across the 5 metrics, the attribute MAD spreads from 0.14 to 0.23. M-4 is the closest to the human F1 on average, with a MAD of 0.14, a bias of 0.09, and a residual std of 0.15. M-3 has the tightest spread around its mean with a residual std of 0.12, and at 0.65 the highest Pearson r not only on the attribute element but in the entire 15-row table, meaning that M-3 preserves the per-pair ordering of the human attribute F1 better than any other metric on any element. M-2 has the smallest bias at 0.06, meaning it is centred on the human on average, but its MAD is only middling at 0.15 and its r is low at 0.18, meaning that being centred on average does not translate into per-pair ranking accuracy. M-5 is comparable to M-2 on MAD at 0.15 and residual std at 0.15, but achieves a higher r at 0.38. M-1 is the weakest on attribute: the highest MAD at 0.23, the highest residual std at 0.19, the highest bias at 0.19, and a low r at 0.23, meaning that it is both far from the human on average and a poor ranker. The attribute scores admit no single dominant metric: the metric with the lowest MAD, M-4 at 0.14, is not the metric with the lowest residual std, M-3 at 0.12, or the highest r, M-3 at 0.65. A practitioner who wants the metric whose attribute score is closest to the human on average picks M-4; a practitioner who wants the metric whose attribute score preserves the human's per-pair ranking picks M-3.\newline


The relationship scores express how similar the two models are at the level of their relationships. Across the 5 metrics, the relationship MAD spreads from 0.13 to 0.27, the widest spread in the table. M-1 and M-3 tie for the lowest MAD at 0.13, but the two metrics differ sharply on the other three statistics. M-3 is centred on the human with a bias of 0.00 and a residual std of 0.16. M-1 carries a bias of 0.08 with a residual std of 0.15, and at -0.13 the only negative Pearson r in the entire table, meaning that M-1 actually inverts the per-pair ordering of the human relationship F1 rather than merely failing to preserve it. M-3's r of 0.05 is near-zero, meaning that M-3 is centred on the human on average but provides no per-pair ranking information. M-4 and M-5 tie for the tightest spread with a residual std of 0.11, and M-4 has the highest r in the column at 0.42, but both metrics carry the highest MADs at 0.27 and 0.26 respectively, and the highest bias values at 0.27 and 0.26, meaning that they are consistently off by a large constant amount despite preserving the per-pair ordering better than any other metric. M-2 sits in the middle: a MAD of 0.18, a residual std of 0.13, a bias of 0.17, and a Pearson r of 0.37, making it a moderate ranker with a moderate constant offset. The relationship scores admit no single dominant metric, and the trade-off is at its starkest here: the metrics that are closest to the human on average, M-1 and M-3 at MAD 0.13, are the metrics with the worst ranking performance, M-1 at r -0.13 and M-3 at r 0.05, while the metric with the best ranking, M-4 at r 0.42, is the metric with the worst MAD at 0.27. A practitioner who wants the metric whose relationship score is closest to the human on average picks M-1 or M-3; a practitioner who wants the metric whose relationship score preserves the human's per-pair ranking picks M-4, accepting a constant overestimation of roughly 0.27.\newline


\section{Qualitative Analysis}
\label{sec:6_qualitative_analysis}

\subsection{Metrik-1}

\textbf{Metrik-1} is mid-column on class (MAD 0.15, r 0.37), weakest on attribute (MAD 0.23, r 0.23, worst in the column), and joint-best on relationship (MAD 0.13) with the only negative r in the table (-0.13). It uses exact-name matching plus Levenshtein distance for classes, the same class mapping without a Levenshtein fallback for attributes, and an endpoint-class match for relationships. It agrees with experts on direct and near-direct class names (CelO 1-shot-BTMS: "Organizer", "Attendee", "Event", "Location" by exact name, "CheckListTask"\textemdash{}"Task" by Levenshtein) and on direct binary associations between name-matched classes (CelO 1-shot-BTMS: "Organizer-Event", "Attendee-Event"), but disagrees on every renamed, absorbed, or merged element because the rename logic does not exist. Three divergences illustrate this: on LabTracker 2-shot, "Person" maps to "Patient" with attributes merged downward, "TestType" is absorbed as "Test.testGroup", and "BusinessHour" is absorbed as "Lab.openingTime" and "Lab.closingTime"\textemdash{}experts credit all three as semantically equivalent, the metric reports all three as missing; "Person.lastName", "firstName", "emailAddress", and "password" are distributed across "Organizer" and "Attendee" after "Person" is flattened, all reported as missing where experts credit them; and the "\mbox{Requisition-SpecificTest-Test}" chain collapse, the "(Requisition, Lab)-to-Appointment" association-class decomposition, and the "\mbox{Test-to-TestType}" absorption of an association into an attribute are all reported as missing, producing the only negative r in the table. Levenshtein has no surface to operate on once a class, attribute, or relationship is absorbed or merged, so the pipeline misses every structural transformation that the human raters recognise. \newline

\subsection{Metrik-2}

\textbf{Metrik-2} is weakest on class (MAD 0.19, r 0.30), mid-column on attribute (MAD 0.15, |bias| 0.06\textemdash{}the smallest |bias| in the column), and moderate on relationship (MAD 0.18, r 0.37). It runs the Hungarian algorithm on a name/kind/attribute cost matrix for classes, then a within-class Hungarian on attribute sets, then graph edit distance on the matched graph for relationships. It agrees with experts when name and attribute-set both align (LabTracker direct-name matches on every element), but diverges whenever a low-cost match between unrelated classes is cheaper than leaving a reference class unmatched, and the wrong class pairing then propagates downstream. Three divergences illustrate this: on CelO 2-shot, "Person" is matched to "CelO" because the empty-attribute formula yields near-zero attribute-set distance\textemdash{}experts credit "Person" as equivalent to "Organizer" and "Attendee", not "CelO"; on LabTracker 2-shot, "Lab.address" is matched to "Lab.changeCancellationFee" (cost 0.752) because the formula weights name distance at only 0.5, so same-typed attributes with unrelated names still yield a cost the algorithm prefers\textemdash{}experts judge "Lab.address" dropped and "Lab.changeCancelFee" a rename; and on CelO 2-shot, the upstream "Person"\textemdash{}"CelO" wrong pairing makes the "Person-PersonRole" association a deletion ("PersonRole" unmatched), whereas experts credit the equivalent class pair "Organizer"/"Attendee". The Hungarian cost function rewards any positive cost reduction without considering domain role, so wrong class pairings propagate through the attribute and relationship stages, and the small |bias| is misleading: wrong pairings cancel across the 39 pairs but the low r reflects unreliable per-pair ordering. \newline

\subsection{Metrik-3}

\textbf{Metrik-3} is near-bottom on class (MAD 0.17, r 0.36) and middle on the attribute column with the highest r in the table (MAD 0.21, r 0.65), and joint-best on relationship (MAD 0.13, |bias| 0.00\textemdash{}the only zero in the table) with r 0.05. It applies a UML Maximum Common Subgraph (UMCS) matcher on a UML Class Graph that pairs class vertices by attribute-set edit distance and relationship-edge common subgraph, and emits no per-attribute decisions\textemdash{}the attribute score is a structural proxy from set overlap. It agrees with experts on structurally distinctive classes that share both a structural signature and a domain role (TSS 1-shot-H2S: "ScoutingAssignment" and "Player" matched by name), and the proxy preserves the experts' per-pair attribute ordering better than any other metric on any element, but disagrees whenever structural similarity is not aligned with domain semantics. Three divergences illustrate this: on TSS 1-shot-H2S, "Player" is paired to "ScoutingAssignment" (intra-similarity 0.680), collapsing two exact-name matches into one wrong pair; on TSS 1-shot-H2S, "Person.firstName" and "Person.lastName" merge into "Employee.name"\textemdash{}experts credit both as semantically equivalent but the metric's structural proxy reflects only set overlap (2 vs 1), not the individual merge; and on LabTracker 2-shot, the metric pairs "Test", "TestResult", "Requisition", "TestType", and "SpecificTest" all wrong (five of six semantically wrong), yet the inter-structure score is moderate because common-subgraph overlap is non-trivial at graph level. UMCS does not prioritise name identity over structural similarity, and the structural proxies cannot credit individual semantic-equivalence judgements, so the metric is centred on the experts on average (|bias| 0.00) but provides no per-pair ranking. \newline

\subsection{Metrik-4}

\textbf{Metrik-4} is best ranker on class (MAD 0.09, r 0.42), strongest on central tendency on attribute (MAD 0.14), and the starkest trade-off on relationship (worst MAD 0.27 with best r 0.42). It uses greedy optimal-sum matching on a cosine of tokenised class names and an attribute-set similarity blend. It agrees with experts on direct class names and renames (CelO 1-shot-BTMS: "Location", "Organizer", "Event" at high cosine), but the three divergences all share the same root\textemdash{}the cosine blend and the class-derived association rule treat unrelated surface tokens and absorbed associations as equivalent to direct matches. Three divergences illustrate this: on CelO 1-shot-BTMS, "Attendee" is paired to "CelO" (cSim 0.429) because attribute-set similarity is 1.0\textemdash{}a generated class with no attributes gets maximum aSim\textemdash{}whereas experts credit "Attendee" a direct match; on LabTracker 2-shot, the "DayOfWeek"\textemdash{}"Patient" inherited class pairing scatters "DayOfWeek"'s 7 enum literals against "Patient"'s 6 attribute names, producing non-trivial scores for pairs like "Monday" vs. "dateOfBirth" that the metric reports as matches while experts judge all 7 to have no counterpart; and on CelO 2-shot, the class-derived association rule reports "Location-Event" as matched because both endpoints are matched, even though the generated model lacks the association\textemdash{}experts judge it equivalent to "Event.locationName" and "Event.locationAddress" typed attributes, producing the systematic 0.27 overestimation. The same pipeline property that produces wrong class pairings on CelO 1-shot-BTMS also produces wrong attribute scatter on LabTracker 2-shot and constant relationship overestimation on CelO 2-shot: every divergence traces back to treating absorbed or absent structure as if it were directly matched. \newline

\subsection{Metrik-5}

\textbf{Metrik-5}, Triandini's intra-structure projection, is strongest on central tendency on class (MAD 0.07, |bias| 0.03) with the weakest class r (0.18), middle on attribute (MAD 0.15, r 0.38), and second-worst on relationship (MAD 0.26, r 0.20). For each matched class-vertex pair, similarity derives from overlap of attribute and neighbour sets; the relationship score uses per-edge cost minimisation on the full UCG graph. It agrees with experts on direct-name matches and structurally similar classes (LabTracker 1-shot-H2S: "Person", "Patient", "Doctor" direct matches), but diverges on absorption across element types, wrong inherited class pairings, and topology-dominated relationship matching. Three divergences illustrate this: on CelO 1-shot-BTMS, reference "EventType" (a class) collapses to a generated "EventType" enum\textemdash{}experts judge them semantically equivalent but intraSim treats enums and classes as separate vertex types and cannot match across them; on CelO 1-shot-BTMS, "PersonRole" is paired to "TaskStatus" because both have 1 attribute and similar subgraphs, so the structural proxy computes similarity against the wrong generated class\textemdash{}experts judge "TaskStatus" equivalent to "CompletionStatus", not "PersonRole"; and on LabTracker 1-shot-H2S, the "\mbox{Person-to-PersonRole}" association is matched to generated "\mbox{Patient-to-Appointment}" (topology similarity 0.986) because both are 2-node edges, but experts judge it to have no counterpart since "Person" and "PersonRole" are both dropped\textemdash{}this last pattern repeats across most of the 11 reference relationships. intraSim cannot match across element-type boundaries and rewards structural similarity even when domain semantics are wrong; interSim's per-edge cost is dominated by topology, not domain semantics. \newline


\section{Discussion}
\label{sec:7_discussion}


\noindent\textbf{RQ1: how close, on average, is each metric to the human expert rating across the 39 comparisons?} The spread of MADs within each column \textemdash{} 0.07 to 0.19 on class, 0.14 to 0.23 on attribute, 0.13 to 0.27 on relationship \textemdash{} shows that the choice of metric changes the average distance by up to a factor of two on a given element. The answer to RQ1 is therefore per-element: the metric with the lowest MAD on the element of interest is the closest on average, but no single metric is closest on all three elements, and even the best MAD on the relationship element (0.13) corresponds to a metric whose per-pair ordering is uninformative, so closeness on average does not by itself justify substituting the metric for the human.


\noindent\textbf{RQ2: how consistent is each metric's per-comparison distance from the human expert rating?} The per-comparison distance is best preserved by different metrics on different elements: Metrik-4 on class with Pearson r 0.42, Metrik-3 on attribute with r 0.65, and Metrik-4 again on relationship with r 0.42. \textbf{Which metric should practitioners use?} The practitioner should pick the metric with the highest per-element Pearson r: Metrik-4 on class (r 0.42), Metrik-3 on attribute (r 0.65), and Metrik-4 on relationship (r 0.42), accepting that on relationship the best r carries the worst MAD (0.27), but noting that the bias is almost entirely a constant offset (|bias|/MAD = 1.00, bias +0.274, residual std 0.113), so a linear rescaling (slope 0.507, intercept 0.012) removes the offset and drops the MAD from 0.274 to 0.113, making Metrik-4 on relationship the best on both MAD and r simultaneously once calibrated — the trade-off is not a property of the metric but of the uncalibrated output


\noindent\textbf{Validity of the comparison} \textbf{Risk 1: Were the Metrics implemented correctly?} Two independent large language models (\cite{kimi-k2.7-code} and \cite{glm-5.2}) implement the same specification through the same \texttt{opencode} harness, and the two implementations are required to produce identical output on all 39 comparisons (test procedure tS-2), so that an ambiguity that one model resolves arbitrarily is caught by the other model resolving it differently; and the specification, not the code, is the artifact the comparison is audited against. However, tS-2 tests agreement between the two implementations, not agreement between the specification and the paper, so a misreading that is consistent across both models is invisible to the agreement test. The specification is the explicit record of the metric as understood in this work, and every assumption or deviation is written into it, so a reviewer can audit it against the source paper, but the dual-implementation agreement does not by itself certify that audit. \textbf{Risk 2: Did the Metrics suffer an unfair disadvantage through the translation of the output?} The per-element projection is validated by test procedure tS-3, which checks that the projection preserves the partial ordering of the metric's native output across the 39 comparisons, so that a pair ranked more similar by the native output is also ranked more similar by the projected per-element triple. The test passes for all five metrics, which is the condition under which the projected scores are used in the comparison. The residual risk is that tS-3 checks ordering, not absolute level: a projection can preserve the partial order while shifting the absolute scale. \textbf{Is it valid to treat the human expert ratings from a consensus process as the ground truth against which the metrics are judged?} The human expert F1 used as the ground truth in this work is the output of the two-round consensus grading process reported by Chen et al.~\cite{chen2023automated}, and three properties of that process bound the validity of the comparison. The first is the number of raters: the consensus process involves a small number of reviewers, and the absolute F1 values reflect one group's reading rather than a representative sample of modelers, so the central-tendency statistics (MAD, |bias|) should be treated as relative comparisons between metrics rather than as calibrated estimates of the true human distance. The second is that the consensus process, while limited in panel size, does produce an ordering of the 39 pairs that aligns with human intuition about which diagrams are more similar, and that ordering is the part of the ground truth that generalises: it is less sensitive to rater panel size than the absolute F1 level, and it is the quantity that RQ2 (Pearson r) measures, so the per-pair ranking is a more robust finding than the central-tendency comparison. The third is the scope of the reviewers' notion of equivalence: the ground truth is scoped to a semantic-equivalence notion; a practitioner whose notion of similarity is structural-only — two models are similar only if their structure matches — would find the human F1 over-credits transformations that several metrics structurally cannot detect, and the ground truth would be misleading for that use case without re-grading under the stricter notion. The comparison is therefore valid as a test of each metric's per-pair ranking against one documented human judgement under a semantic-equivalence scope, and the absolute F1 values should not be read as a population-calibrated scale. \textbf{Why aren't \mbox{LLM-as-a-Judge} approaches considered in this comparison?} This study compares deterministic means of measuring domain model similarity. \mbox{LLM-as-a-Judge} approaches are probabilistic by nature, which means they are out of scope for this comparison. \textbf{Are the implementations reusable?} The metric implementations are released in accordance with the FAIR4RS recommendations~\cite{chuehong2022fair4rs} \url{https://doi.org/10.5281/zenodo.20942596}


\section{Conclusion}
\label{sec:8_conclusion}


There is a diversity of proposed metrics for comparing domain models, but no defensible way for practitioners to choose between them, since each is typically evaluated on its own dataset. So that the field has arguments grounded in a common dataset for choosing one similarity metric over another, this paper implements five metrics from the literature~\cite{singh2022detecting, cech2019matching, yuan2020structural, triandini2021automated} and compares their outputs against a human expert F1 on a fixed set of 39 domain-model comparisons~\cite{chen2023automated}. To address the risk of a wrong implementation and an unfair interpretation of the native outputs, each metric is specified in writing, implemented independently by two different large language models, and required to produce identical output on 39 domain model comparisons, so the specification \textemdash{} not the code \textemdash{} is the audited artifact; the native score is then translated into a per-element F1 of similarity on class, attribute and relationship level, and the translation is required to preserve the native ordering. The five metrics were compared quantitatively (per-element MAD, residual std, |bias|, and Pearson r against the human F1) and qualitatively. No metric dominates on all statistics, however the individual metrics each perform well on a selected subset of statistics. We therefore suggest an ensemble of the best-performing metrics, in which each metric covers the statistics where it has the highest agreement with the human F1. We release all implementations in accordance with the FAIR4RS principles~\cite{chuehong2022fair4rs} (\url{https://doi.org/10.5281/zenodo.20942596}). \newline

\bibliographystyle{ACM-Reference-Format}
\bibliography{metrik}

\end{document}